# The Structure of the Multiverse

## David Deutsch

**Centre for Quantum Computation**
**The Clarendon Laboratory**
**University of Oxford, Oxford OX1 3PU, UK**

*The structure of the multiverse is determined by information flow.*

## 1. Introduction

The idea that quantum theory is a true description of physical reality led Everett
(1957) and many subsequent investigators (e.g. DeWitt and Graham 1973, Deutsch
1985, 1997) to explain quantum-mechanical phenomena in terms of the simultaneous
existence of parallel universes or histories. Similarly I and others have explained the
power of quantum computation in terms of 'quantum parallelism' (many classical
computations occurring in parallel). However, if reality – which in this context is
called the *multiverse* – is indeed literally quantum-mechanical, then it must have a
great deal more structure than merely a collection of entities each resembling the
universe of classical physics. For one thing, elements of such a collection would
indeed be 'parallel': they would have no effect on each other, and would therefore
not exhibit quantum interference. For another, a 'universe' is a global construct –
say, the whole of space and its contents at a given time – but since quantum
interactions are local, it must in the first instance be local physical systems, such as
qubits, measuring instruments and observers, that are split into multiple copies, and
this multiplicity must propagate across the multiverse at subluminal speeds. And for
another, the Hilbert space structure of quantum states provides an infinity of ways of slicing up the multiverse into 'universes', each way corresponding to a choice of basis. This is reminiscent of the infinity of ways in which one can slice ('foliate') a spacetime into spacelike hypersurfaces in the general theory of relativity. Given such a foliation, the theory partitions physical quantities into those 'within' each of the hypersurfaces and those that relate hypersurfaces to each other. In this paper I shall sketch a somewhat analogous theory for a model of the multiverse.

The quantum theory of computation is useful in this investigation because, as we shall see, the structure of the multiverse is determined by information flow, and the universality of computation ensures that by studying quantum computational networks it is possible to obtain results about information flow that must also hold for quantum systems in general. This approach was used by Deutsch and Hayden (2000) to analyse information flow in the presence of entanglement. In that analysis, as in this one, no quantitative definition of information is required; the following two qualitative properties suffice:

- **Property 1**: A physical system $S$ *contains information* about a parameter $b$ if (though not necessarily only if) the probability of some outcome of some measurement on $S$ alone depends on $b$.

- **Property 2**: A physical system $S$ *contains no information* about $b$ if (and for present purposes we need not take a position about 'only if') there exists a complete description of $S$ that is independent of $b$.

I shall assume that an entity $S$ qualifies as a 'physical system' if (but not necessarily only if) it is possible to store information in $S$ and later to retrieve it. That is to say, it must be possible to cause $S$ to satisfy the condition of Property 1 for containing information about some parameter $b$. It is implicit in this, and in Properties 1 and 2, that $b$ must be capable of taking more than one possible value, so there must exist some suitable sense in which if $S$ contained different information it would still be the same physical system. This condition raises interesting questions about the counter-factual nature of information which it will not be necessary to address here. It is also necessary that $S$ be identifiable as the same system over time. This is particularly straightforward if $S$ is causally autonomous – that is to say, if its evolution depends on nothing outside itself.

## 2. Classical computers

Consider a classical reversible computational network containing $N$ bits $B_1 \ldots B_N$. A specification of the values $b_1(t), \ldots, b_N(t)$ of the bits just after the $t$'th computational step constitutes a complete description of the computational state of the network at that instant. Given the structure of the network (its gates, and how the carriers of the bits move between them), this also determines the computational state just after every other computational step. We are not interested in the network's state *during* computational steps, nor in its non-computational degrees of freedom, because we know that the computational degrees of freedom at integer values of $t$ form a causally autonomous system, and it is that system which we shall regard as faithfully modelling, with some finite but arbitrarily high degree of accuracy, the flow of information in a classical system or classical universe.

Information flow in the network is *local* in the sense that if some information is confined to a set of bits $C$ at time $t$, then at time $t+1$ that information is confined to bits that have passed through the same gate as some member of $C$ during the $(t+1)$'th computational step. In particular, if a network consists of two or more sub-networks that are disconnected for a period, then information cannot flow from one of those sub-networks to another during that period. Where a system $S$ has local dynamics – for instance, if it is a field governed by a differential equation of motion – and we want to draw conclusions about information flow in $S$ by studying networks that model $S$ to some degree of approximation, we must consider only models with the property that local regions of $S$ correspond to local (in the above sense) regions of the network.

If we were to construct such a network in the laboratory, then each of the $2^N$ possible bit-sequences $b_1, \ldots, b_N$ would specify a physically and computationally different state of the network. But if reality *consisted* of such a network, that would not necessarily be so, because there would then be no external labels, such as spatial location, to distinguish one bit from another. So, for instance, if the network consisted of two disjoint sub-networks with identical structures, containing bits $B_1, \ldots, B_{N/2}$ and $B_{N/2+1}, \ldots, B_N$ respectively, then any two bit-sequences of the form $_1, \ldots, _N$ and $_{N/2+1}, \ldots, _N, _1, \ldots, _{N/2}$ would refer to the same physical state. The same applies when we are considering a hypothetical network that models information flow in reality as a whole: if the structure of such network is invariant under some permutation    of its bits, then any two bit-sequences that are related by    refer to the same state of reality.

Let us refer to a bit-sequence $b_1(t), \ldots, b_N(t)$ collectively as $b(t)$ (which can be thought of as the binary number $2^{N-1} b_N(t) + \ldots + 2 b_2(t) + b_1(t)$   $Z_{2^N}$). During each computational step, the values of the bits in the network change according to

$$b(t+1) = f_t(b(t)), \tag{1}$$

where each $f_t$ is some invertible function from $Z_{2^N}$ to itself, which characterises the action of all the gates through which the bits pass during the $(t+1)$'th computational step.

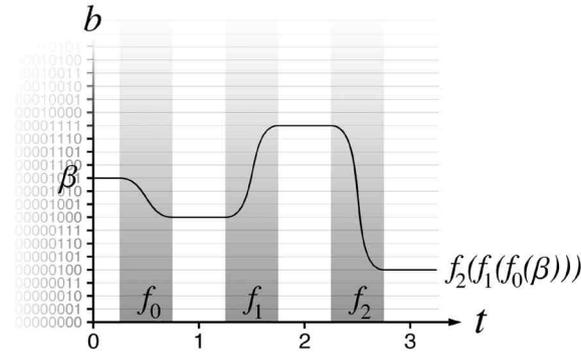

Fig. 1: History of a classical computation

The course of such a computation with initial state $b(0) = \beta$ is shown schematically in Fig. 1. The parts of the graph in the shaded regions (*i.e.* during computational steps), and the non-integer values of $b$, have no significance except to indicate that the motion of a real computer would interpolate smoothly between computational states.

## 3.  Ensembles of classical computers

Consider a collection of $M$ classical networks of the kind described in Section 2, all with the same structure in terms of gates, but not necessarily all starting in the same initial state. One way of describing such a collection is as a single network consisting of $M$ disconnected sub-networks. The network has $NM$ bits $B_1 \ldots B_{NM}$ , where $B_1 \ldots B_N$ belong to the 'first' sub-network, $B_{N+1} \ldots B_{2N}$ to the 'second', and so on. But since the structure of the network is invariant under any permutation of the sub-networks, we must regard any pair of bit-sequences of length $NM$ that are related by such a permutation as referring to physically identical states.

In other words, when such sub-networks are in identical states, they are *fungible*. The term is borrowed from law, where it refers to objects, such as banknotes, that are deemed identical for the purpose of meeting legal obligations. In physics we may define entities as fungible if they are not merely deemed identical but *are* identical, in the sense that although they can be present in a physical system in varying numbers or amounts, permuting them does not change the physical state of that system.

Fungibility is not new to physics. Many physical entities, such as amounts of energy, are fungible even in classical physics: one can add a Joule of energy to a physical system, but one cannot later extract the same Joule. In quantum physics some material objects – bosons – are fungible too: it makes sense to ask how many identical photons there are in a cavity, and it makes sense to add one more of the same kind and then to remove one, but it does not make sense to ask whether the photon that has been removed is or is not the photon that was previously added (unless there was exactly one photon of that kind present).

Hence an alternative way of describing our $NM$-bit network is as a *multiset* of $M$ networks, each with $N$ bits. A multiset is like a set except that some of its elements are fungible. Each element is associated with an integer, its *multiplicity*, which specifies how many instances of it appear in the multiset. In the present case, if $\mu_b(t)$ is the number of sub-networks that are in the state $b \in Z_{2^N}$ at time $t$, then the state of the network at time $t$ is completely specified by the $2^N$ multiplicities $\{\mu_b(t)\}$. Their equation of motion is

$$\mu_b(t+1) = \mu_{f_t^{-1}(b)}(t). \tag{2}$$

An ensemble is a limiting case of a multiset where the total number of elements $M$ goes to infinity but all the proportions $\mu_b(t)/M$ tend to definite limits. Considered as a function of $b$, $\lim_M(\mu_b(t)/M)$ is the distribution function for the ensemble at time $t$. Henceforth in this paper I shall use the term 'ensemble' for both ensembles and multisets, and the term 'multiplicity' to denote both discrete multiplicities and real-valued proportions.

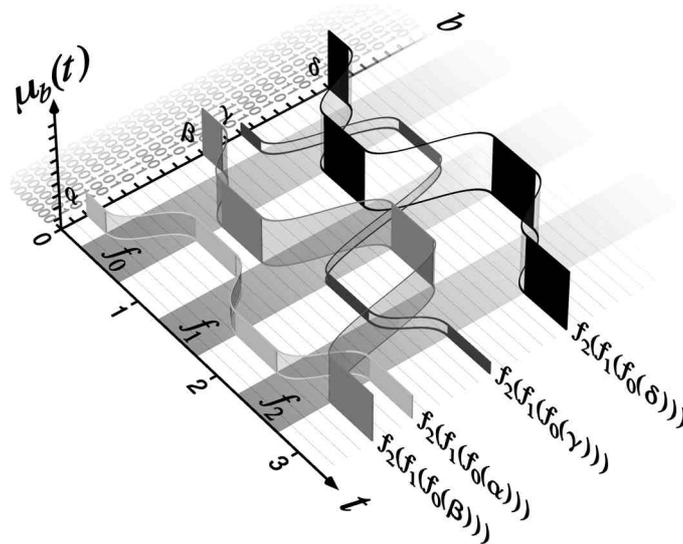

**Fig. 2: History of an ensemble of reversible classical computations**

Fig. 2 illustrates schematically a computation being performed by an ensemble of twelve computers. Four of them are performing the same computation as the computer referred to by Fig. 1, with input . Three other computations are being performed in parallel with that one. They have inputs , and , and are being performed by two, one and five of the computers respectively.

I shall refer to a non-empty sub-ensemble in which all the computers are in the same state as a *branch* of the ensemble – so for instance, the ensemble in Fig. 2 has four branches throughout the computation. Note the following elementary properties of branches under reversible classical physics. First, the total number of branches is conserved: they cannot split, join, come into existence or be destroyed. Second, the multiplicity of each branch is conserved. And third, none of the branches affect each other; that is to say, the behaviour of each branch is determined by its own initial state and $f$, and is therefore independent of how many other branches are present and what their states and multiplicities are. These properties give each branch a well-defined identity over time, even though the values of its bits change.

There are such things as fungible processes as well as fungible objects. Because each of the computations takes place within a particular branch over time, and the whole ensemble over time is invariant under permutations of the computations within a branch, those computations are themselves fungible.

In this representation of our network as an ensemble $E$, the equation of motion (2) specifies how the multiplicity of a fixed bit-sequence $b$ changes with time. This is not well suited to the analysis of information flow because, as illustrated in Fig. 2, the information in $E$ flows entirely in branches which are characterised by *constant* multiplicities and *time-varying* bit-sequences. It is possible to make this manifest by using an alternative representation of an ensemble of classical reversible systems that bears the same relationship to the standard one as the Heisenberg picture does to the Schrödinger picture in quantum theory. To motivate the representation, consider first a list $b(0)$ of the computational states of all the branches in the ensemble at time zero, in any order. For example, $b(0)$ for the ensemble of Fig. 2 could be $(\ ,\ ,\ ,\ )$. It will turn out to be convenient to consider such lists to be vectors, with an algebra that I shall define below, and I shall call such vectors *e-numbers* by analogy with the terms 'q-number' for quantum operators and 'c-number' for scalars. For each $t > 0$, define the e-number $b(t)$ as a similar list, or vector, of the states of all the branches at time $t$, with the branches appearing in the same order as they do in $b(0)$. So in the Fig. 2 case, $b(1) = \left(f_0(\ ), f_0(\ ), f_0(\ ), f_0(\ )\right)$, and so on. Thus each component of $b(t)$, as $t$ varies, is the evolving state of one particular branch of the ensemble. Since the multiplicities of branches are constant, we can list them as a single, constant e-number $\mu = \left(\mu\ (0), \mu\ (0), \mu\ (0), \mu\ (0)\right)$.

The quantities $\mu$ and $b(t)$ together contain the same information as the $\{\mu_b(t)\}$, and therefore amount to a complete description of the state of the ensemble at time $t$.

In general, e-numbers for an ensemble of $N$-bit reversible classical networks may be defined as follows. They are elements of a $2^N$-dimensional vector space $V$. (Lower-dimensional representations are sometimes possible, as in the example above.) We can express $b(t)$, or any other e-number, in terms of an orthonormal basis $\{P_b(t)\}$

$$b(t) = \sum_b b P_b(t). \tag{3}$$

A straightforward representation for $P_b(0)$ would be a list of length $2^N$, whose $(b+1)$'th element was 1, with the rest all 0. The equation of motion for the ensemble is

$$P_b(t+1) = P_{f_t^{-1}(b)}(t). \tag{4}$$

I shall refer to $P_b(t)$ as the 'projector for the bits to take the values $b$ at time $t$'. The reason for this terminology is, as will become apparent below, that these are the e-number analogues of quantum projection operators or Boolean observables. Note that they are not projection operators on $V$, but elements of $V$.

If $\{\lambda_b\}$ are any c-number coefficients and $g$ is any function initially defined on c-numbers, we can define

$$g\left(\sum_b \lambda_b P_b(t)\right) \equiv \sum_b g(\lambda_b) P_b(t). \tag{5}$$

From (3), (5) and (4) it follows that

$$b(t+1) = f_t\Big(b(t)\Big). \tag{6}$$

Hence the equation of motion (6) of the e-bits has the same form as its counterpart (1) for a single classical computer, with e-numbers replacing integers.

Apart from multiplication by a c-number in the usual vector-space manner, the algebra has three further forms of multiplication, which may all be defined by their actions on the projectors: The scalar product $x.y$ is defined by

$$P_a(t).P_b(t) = \quad_{ab}. \tag{7}$$

The e-number product $xy$, which is the e-number analogue of the classical or quantum product of observables, is defined by

$$P_a(t)P_b(t) = P_a(t)\quad_{ab}. \tag{8}$$

We can also define a unit e-number $\mathbf{1} = \quad_b P_b(t)$ for any t, and a zero e-number $\mathbf{0} = 0$, which have the usual multiplicative properties with respect to e-number multiplication. The tensor product $x \quad y$, which is used for combining ensembles, could, for instance, be defined by

$$P_a^E(t) \quad P_b^{E'}(t) = P_{2^N a + b}^{E \times E'}(t), \tag{9}$$

where the two projectors on the left of (9) refer to ensembles $E$ and $E'$, the second having $N$ bits per element, and the projector on the right refers to the combination $E \times E'$ of those two ensembles.

The connection with the conventional representation is

$$\mu_b(t) = \mu.P_b(t). \tag{10}$$

We can now regard Fig. 2 as showing the history of $b(t)$ in the 'state' $\mu$, rather than the history of the $\{\mu_b(t)\}$. Note that from a complete specification of the algebra generated by the e-numbers $\mu$, $b(t)$ and $\mathbf{1}$, (i.e. a means of calculating the scalar products of all the expressions that can constructed from those e-numbers by addition, scalar multiplication and e-number multiplication), one can obtain the projectors $P_b(t)$ for all states $b$ that are present in the ensemble at any time $t$:

$$P_b(t) = \quad \Big(b(t) - b\mathbf{1}\Big), \tag{11}$$

where    is an e-number version of the Krönecker delta function, defined by

$$\langle X \rangle = \frac{1}{(2^N - 1)!} \prod_{b=1}^{2^N - 1} (b\mathbf{1} - X). \tag{12}$$

It then follows from (10) that a complete description of the ensemble is contained entirely in the algebra of its e-number descriptors $\mu$, $b(t)$ and $\mathbf{1}$, independently of any particular representation of these e-numbers as $2^N$-tuples.

## 4. Quantum computers performing classical computations

The central question addressed in this paper can now be stated as follows: in what sense, and in what approximation, can a quantum computation be said to contain an ensemble of classical computations?

Consider a quantum computational network containing $N$ qubits $Q_1, \ldots, Q_N$. Following Gottesman (1999) and Deutsch and Hayden (2000), let us represent each qubit $Q_k$ at time $t$ in the Heisenberg picture by a triple

$$\hat{\mathbf{b}}_k(t) = \left( \hat{b}_{kx}(t), \hat{b}_{ky}(t), \hat{b}_{kz}(t) \right) \tag{13}$$

of $2^N \times 2^N$ Hermitian matrices representing Boolean observables (projection operators) of $Q_k$, satisfying

$$\left[ \hat{\mathbf{b}}_k(t), \hat{\mathbf{b}}_{k'}(t) \right] = 0 \qquad\qquad (k \neq k')$$

$$\left( \hat{\mathbf{1}} - 2\hat{b}_{kx}(t) \right)\left( \hat{\mathbf{1}} - 2\hat{b}_{ky}(t) \right) = i\left( \hat{\mathbf{1}} - 2\hat{b}_{kz}(t) \right)$$

$$\hat{b}_{kx}(t)^2 = \hat{b}_{kx}(t) \qquad\qquad \text{(and cyclic permutations over } (x, y, z)) \tag{14}$$

The Heisenberg state $| \ \rangle$ of the network is a constant, so we can adopt the abbreviated notation $\langle \hat{X} \rangle \equiv \langle \ | \hat{X} | \ \rangle$ for the expectation value of any observable $\hat{X}$ of the network.

The effect of an *n*-qubit quantum gate during one computational step is to transform the *3n* matrices representing the *n* participating qubits into functions of each other in such a way that the relations (14) are preserved.

Rotations of any descriptor $\hat{\boldsymbol{b}}_k(t)$, considered as a (matrix-valued) 3-vector in the Euclidean *x*-*y*-*z* space, are such functions. This allows a large class of possible alternative representations of the qubits, corresponding to the freedom to make such rotations for each qubit independently and then to redefine all the *x*-, *y*- and *z*-directions. By convention we use this freedom to choose a representation (if one is available) in which the *z*-components $\hat{b}_{kz}(t)$ are stabilised by any decoherence or measurement that may occur, or more generally, in which those components are performing classical computations (see below). In any case, we can define

$$\hat{b}(t) = 2^{N-1}\hat{b}_{Nz}(t) + \ldots + 2\hat{b}_{2z}(t) + \hat{b}_{1z}(t), \tag{15}$$

as for a classical computer, though note that $\hat{b}(t)$ is not a complete specification of the state of the quantum computer at time *t*: there are also the other components of the descriptors $\left\{\hat{\boldsymbol{b}}_k(t)\right\}$, and the Heisenberg state $|\ \rangle$. In principle, one could change the representation at every computational step, but that adds no generality, being the same as studying a different network in a constant representation. It would also be possible to construct alternative representations that were related to this one by more general transformations that are not expressible as compositions of single-qubit transformations. However, these would not be appropriate in the present investigation because the 'qubits' in such representations would not be local in the network, and in order to model information flow we are using local interactions (gates) of the network to model local interactions in general quantum systems.

A quantum network (or sub-network) is said to be 'performing a classical computation' during the $(t+1)$'th computational step if its $\hat{b}(t+1) = f\left(\hat{b}(t)\right)$ for some function *f* (not necessarily invertible). This occurs if and only if all its gates that act on qubits during that step have classical analogues – including one-qubit gates with the effect

$$\hat{\boldsymbol{b}}_k(t+1) = \Big(\text{anything}, \quad \text{anything}, \quad \hat{b}_{kz}(t)\Big), \tag{16}$$

which is a non-trivial quantum computation but corresponds to the classical gate whose only computational effect is a one-step delay. That is not to say that the quantum network *is* a classical computer during such a period: it still has qubits rather than bits; it (or at least, the network as a whole) is still undergoing coherent motion; and its computational state is not specified by any sequence of N binary digits.

The Toffoli gate, which is universal for reversible classical computations, is defined as having the following effect on the $k$'th, $l$'th and $m$'th bits of a classical network:

$$\begin{array}{lcl} b_k(t+1) & & b_k(t) \\ b_l(t+1) & = & b_l(t) \\ b_m(t+1) & & b_m(t) + b_k(t)b_l(t) - 2b_k(t)b_l(t)b_m(t) \end{array} . \tag{17}$$

It follows from the results of Section 3 that in an ensemble of networks containing a Toffoli gate, its effect has the same functional form as (17), with e-numbers replacing c-numbers:

$$\begin{array}{lcl} b_k(t+1) & & b_k(t) \\ b_l(t+1) & = & b_l(t) \\ b_m(t+1) & & b_m(t) + b_k(t)b_l(t) - 2b_k(t)b_l(t)b_m(t) \end{array} . \tag{18}$$

Compare this with the effect of the quantum version of the Toffoli gate:

$$\begin{array}{lcl} \hat{\boldsymbol{b}}_k(t+1) & & \Big(\hat{b}_{kx} + \hat{b}_{lz}\hat{b}_{mx} - 2\hat{b}_{kx}\hat{b}_{lz}\hat{b}_{mx}, \quad \hat{b}_{ky} + \hat{b}_{lz}\hat{b}_{mx} - 2\hat{b}_{ky}\hat{b}_{lz}\hat{b}_{mx}, \qquad\qquad \hat{b}_{kz} \qquad\qquad \Big) \\ \hat{\boldsymbol{b}}_l(t+1) & = & \Big(\hat{b}_{lx} + \hat{b}_{kz}\hat{b}_{mx} - 2\hat{b}_{kz}\hat{b}_{lx}\hat{b}_{mx}, \quad \hat{b}_{ly} + \hat{b}_{kz}\hat{b}_{mz} - 2\hat{b}_{kz}\hat{b}_{ly}\hat{b}_{mx}, \qquad\qquad \hat{b}_{lz} \qquad\qquad \Big) \\ \hat{\boldsymbol{b}}_m(t+1) & & \Big(\qquad\qquad \hat{b}_{mx}, \qquad\qquad \hat{b}_{my} + \hat{b}_{kz}\hat{b}_{lz} - 2\hat{b}_{kz}\hat{b}_{lz}\hat{b}_{my}, \quad \hat{b}_{mz} + \hat{b}_{kz}\hat{b}_{lz} - 2\hat{b}_{kz}\hat{b}_{lz}\hat{b}_{mz}\Big) \end{array} \tag{19}$$

For the sake of brevity, the parameter $t$ has been suppressed from all the matrices on the right of (19). It is easily verified that the conditions (14) are preserved by this transformation. Notice that the $z$-components $\hat{b}_{kz}(t+1), \hat{b}_{lz}(t+1), \hat{b}_{mz}(t+1)$ of the descriptors of the qubits emerging from the gate (third column on the right of (19)) depend only on the $z$-components $\hat{b}_{kz}(t), \hat{b}_{lz}(t), \hat{b}_{mz}(t)$ of the descriptors of the qubits entering the gate. Notice also that these $z$-components commute with each other and that their equation of motion has the same functional form as that of the corresponding ensemble of classical computers (18). Given the universality of the Toffoli gate, all these properties must hold whenever a quantum network, or any part of it, performs a classical computation. In other words, whenever any quantum network (including a sub-network of another network) is performing a classical computation $f$, the matrices $\left\{ \hat{b}_{kz}(t) \right\}$ for that network evolve independently of all its other descriptors. Moreover, under the following correspondence

| Ensemble | Quantum | |
|----------|---------|---|
| $b_k(t)$ | $\hat{b}_{kz}(t)$ | |
| $b(t)$ | $\hat{b}(t)$ | |
| $\mu$ | $\mid\;\rangle\langle\;\mid$ | |
| $1$ | $\hat{1}$ | (20) |
| $P_b(t)$ | $\hat{P}_b(t) \;\; \mid b;t\rangle\langle b;t\mid$ | |
| $X.Y$ | $Tr\hat{X}\hat{Y}$ | |
| $XY$ | $\hat{X}\hat{Y}$ | |
| $X \;\; Y$ | $\hat{X} \;\; \hat{Y}$ | |

the commuting algebra of these matrices forms a faithful representation of the algebra of e-numbers describing an ensemble of classical networks performing $f$. In (20), $\mid b;t\rangle$ is the eigenvalue-$b$ eigenstate of $\hat{b}_{kz}(t)$, and $\hat{X}$ and $\hat{Y}$ are the same functions of the $\left\{ \hat{b}_{kz}(t) \right\}$ as $X$ and $Y$ respectively are of the $\left\{ b_k(t) \right\}$.

We also have $\hat{b}(t+1) = f_t\left( \hat{b}(t) \right)$, the analogue of (6). Thus Fig. 2, showing the course of an ensemble of classical computations, could equally well be a graph of the quantities $\left\langle \hat{P}_b(t) \right\rangle$ in a quantum computer that was performing the same classical computation as that ensemble. Note also that while the quantities $\mu . P_b(t)$ form a complete description of the ensemble of classical computations, the $\left\langle \hat{P}_b(t) \right\rangle$ are not a complete description of the state of the quantum computation.

Thus in any sub-network $R$ of a quantum computational network where a reversible classical computation is under way, half the parameters describing $R$ are precisely the descriptors of an ensemble of classical networks. It is half the parameters because, from (14), any two of the three components of $\left\{ \hat{\boldsymbol{b}}_k(t) \right\}$ determine the third. This does not imply that such a subsystem constitutes half the region of the multiverse in which $R$ exists. Proportions in the latter sense – which formally play the role of probabilities under some circumstances, as shown in Deutsch (1999) – are determined by the Heisenberg state as well as the observables, and do not concern us here because the present discussion is not quantitative.

The other half of the parameters, say the $\left\{ \hat{b}_{kx}(t) \right\}$), contain information that is physically present in $R$ (it can affect subsequent measurements performed on $R$ alone) but cannot reach the ensemble (the descriptors of the ensemble being independent of that information). But the reverse is not true: as (19) shows, information can reach the quantum degrees of freedom from the ensemble.

The proposition that parts of the multiverse have the same description as an ensemble with given properties is not quite the same as the proposition that such an ensemble is actually present in those parts of the multiverse, for the description might refer to entities that are not present in addition to those that are. In particular, an ensemble has an alternative interpretation as a *notional* collection, only one member of which is physically real, with the multiplicity of a given branch representing the probability that the properties of that branch were the ones prepared in the real system at the outset, by some stochastic process. However, no such interpretation is possible if the branches affect each other, as they do in general quantum phenomena, and in quantum computations in particular (see Benjamin 2001).

## 5. Quantum computations

When a quantum computational network is performing a general computation, it need not be the case that the descriptors of any part of the network over two or more computational steps constitute a representation of an evolving e-algebra. There need exist no functions $f_t$ and no choice of the 'z-directions' for defining the $\hat{b}_{kz}(t)$ and hence $\hat{b}(t)$, such that $\hat{b}(t+1) = f_t\left(\hat{b}(t)\right)$, so the conditions discussed in Section 3 for branches to have an identity over time need not hold. At each instant $t$, it is still possible to extract a set of numbers $\left\{ P_b(t) \right\}$ from the description of the network at time $t$, and these still constitute a partition of unity, and still indicate which of the eigenvalues $b$ of the observable $\hat{b}(t)$ are present in the multiverse at time $t$ (in the sense that if $\hat{b}(t)$ were measured immediately after time $t$, the possible outcomes would be precisely the values for which $\left\{ P_b(t) \right\} \neq 0$). But although the physical evolution is of course always continuous, there is in general no way of 'connecting up the dots' in a graph of the quantities $\left\{ P_b(t) \right\}$ against $b$ and $t$ that would correctly represent the flow of information. Hence there exists no entity (such as a 'branch' or 'universe'), associated with only one of the values $b$, that can be identified as a physical system over time.

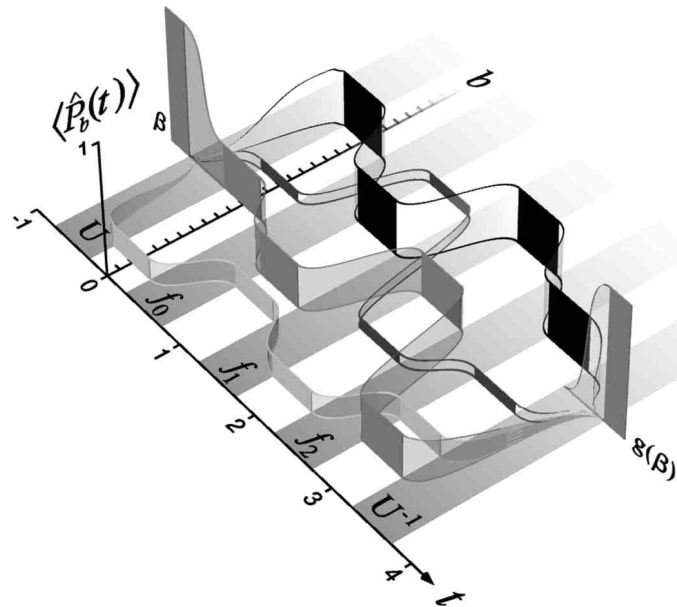

Fig. 3: History of a quantum computation

In a typical quantum algorithm, as illustrated schematically in Fig. 3, the qubits first undergo a non-classical unitary transformation U, then a reversible classical computation, and finally another unitary transformation which is often the inverse $U^{-1}$ of the first one. Despite the fact that the branches lose their separate identities during the periods of the quantum transformations U and $U^{-1}$, we can still track the flow of information reasonably well in terms of ensembles: For $t < -1$, there is a homogeneous ensemble, in all elements of which the computer is prepared with the input . For $-1 < t < 0$, this region of the multiverse does not resemble an ensemble: it has a more complicated structure, but the quantum computer as a whole does still contain the information that the input was . For $0 < t < 3$ an ensemble is present again, this time with four branches. The information about $b$ may no longer be wholly present in that ensemble; some or all of it may be in the other half of the computer's degrees of freedom. For $t > 3$ the story is similar to that for $t < 0$, but in reverse order, so that finally there is a homogeneous ensemble with all elements holding the value $g( )$.

Consider a quantum computation whose $(t+1)$'th step has the effect: 'if qubit $N$ is 1, evaluate the invertible function $f_t$ on qubits 1 to $N-1$, and otherwise perform the unitary transformation $\mathrm{U}_t$ on those qubits'. In other words, during the $(t+1)$'th step the computer performs the transformation:

$$\mathrm{V}_t = \hat{b}_{Nz} \sum_{b=0}^{2^{N-1}-1} | f_t(b) \rangle \langle b | + \left(\hat{1} - \hat{b}_{Nz}\right) \mathrm{U}_t. \qquad (21)$$

on all $N$ qubits. (Since $\hat{b}_{Nz}$ does not change during this process we can drop its parameter $t$.) If the $\mathrm{U}_t$ do not represent classical computations then clearly the network as a whole is not performing a classical computation unless $\left\{\hat{b}_{Nz}\right\} = 1$. Nevertheless, it is still the case that some of the descriptors of this network – only about a quarter of them, this time, namely the $\left\{\hat{b}_{Nz}\hat{b}_{kz}(t)\right\}$ – are those of a causally autonomous ensemble of classical computers, which, by the argument above, means that such an ensemble is present. Half the descriptors, say the $\left\{\left(\hat{1} - \hat{b}_{Nz}\right)\hat{b}_{kz}(t)\right\} \cup \left\{\left(\hat{1} - \hat{b}_{Nz}\right)\hat{b}_{kx}(t)\right\}$, do form a causally autonomous system but do not form a representation of an e-algebra, while the remaining quarter, say, the $\left\{\hat{b}_{Nz}\hat{b}_{kx}(t)\right\}$, are neither causally autonomous nor (therefore) form a representation of an e-algebra. Thus this system has the following information-flow structure: it consists of two subsystems between which information does not flow. One of them is performing a quantum computation and cannot be further analysed into autonomous subsystems; the other contains both an ensemble of $2^{N-1}$ classical computations and a further system that can not be analysed into autonomous subsystems; moreover, information can reach it from the ensemble but not vice-versa.

In this network, the individual branches of the ensemble whose e-number algebra is represented by the matrices $\left\{\hat{b}_{Nz}\hat{b}_{kz}(t)\right\}$, qualify as physical systems according to the criteria of Section 1 because, for instance, if $\hat{X}(t)$ is any observable on the network at time $t$ and $0 \le k < 2^{N-1}$, a measurement of the observable $\hat{b}_{Nz}(t)\hat{P}_k(t)\hat{X}(t)\hat{P}_k(t)\hat{b}_{Nz}(t)$ is a measurement on one such branch alone – the one in which the classical computation is taking place and all the classical computers in the branch are in state $k$ at time $t$.

The 'controlled-not' gate, or measurement gate, which has the effect

$$
\begin{aligned}
\hat{\boldsymbol{b}}_m(t+1) &= \left( \hat{b}_{nx} + \hat{b}_{mx} - 2\hat{b}_{nx}\hat{b}_{mx}, \quad \hat{b}_{nx} + \hat{b}_{my} - 2\hat{b}_{nx}\hat{b}_{my}, \quad \hat{b}_{mz} \right) \\
\hat{\boldsymbol{b}}_n(t+1) &= \left( \hat{b}_{nx}, \quad \hat{b}_{ny} + \hat{b}_{mz} - 2\hat{b}_{ny}\hat{b}_{mz}, \quad \hat{b}_{nz} + \hat{b}_{mz} - 2\hat{b}_{nz}\hat{b}_{mz} \right)
\end{aligned} \quad , \tag{22}
$$

where again the parameter $t$ has been suppressed from all the matrices on the right of the equation, can be used to model the effects of measurement and decoherence. $Q_m$ is known as the 'control' qubit and $Q_n$ the 'target' qubit. Because this is a reversible classical computation (or rather, the quantum analogue of one), the last ($z$-) column of (22) again depends only on the $z$-components $\hat{b}_{mz}(t)$ and $\hat{b}_{nz}(t)$ of the descriptors of the qubits entering the gate. Furthermore, the $z$-component of the descriptor of the control qubit is unaffected by the action of the measurement gate (i.e. $\hat{b}_{mz}(t+1) = \hat{b}_{mz}(t)$). Therefore, if some sub-network of a quantum network performs a classical computation for a period if the network is isolated, and then it is run with some or all of the observables $\left\{ \hat{b}_{kz} \right\}$ being repeatedly measured between computational steps, it will still perform the same classical computation and will contain an ensemble identical to that which it would contain if it were isolated (though its other descriptors will be very different). Since decoherence can be regarded as a process of measurement of a quantum system by its environment, the same conclusion holds in the presence of decoherence. It also holds, by trivial extension, if the classical computation is irreversible, since an irreversible classical computation is simply a reversible classical computation in which some of the information leaves (becomes absent from) the sub-network under consideration.

Since a generic quantum computational network does not perform anything like a classical computation on a substantial proportion of its qubits for many computational steps, it may seem that when we extend the above conclusions to the multiverse at large, we should expect parallelism (ensemble-like systems) to be confined to spatially and temporally small, scattered pockets. The reason why these systems in fact extend over the whole of spacetime with the *exception* of some small regions (such as the interiors of atoms and quantum computers), and why they approximately obey classical laws of physics, is studied in the theory of decoherence (see Zurek 1981, Hartle 1991). For present purposes, note only that although most of the descriptors of physical systems throughout spacetime do not obey anything like classical physics, the ones that do, form a system that, to a good approximation, is not only causally autonomous but can store information for extended periods and carry it over great distances. It is therefore that system which is most easily accessible to our senses – indeed, it includes all the information processing performed by our sense organs and brains. It has the approximate structure of a classical ensemble comprising 'the universe' that we subjectively perceive and participate in, and other 'parallel' universes.

In Section 1 I mentioned that the theory presented here does roughly the same job for the multiverse as the theory of foliation into spacelike hypersurfaces does for spacetime in general relativity. There are strong reasons to believe that this must be more than an analogy. It is implausible that the quantum theory of gravity will involve observables that are functions of a c-number time. Instead, time must be associated with entanglement between clock-like systems and other quantum systems, as in the model constructed by Page and Wootters (1983), in which different times are seen as special cases of different universes. Hence the theory presented here and the classical theory of foliation must in reality be two limiting cases of a single, yet-to-be-discovered theory – the theory of the structure of the multiverse under quantum gravity.

## Acknowledgement

I wish to thank Dr. Simon Benjamin for many conversations in which the ideas leading to this paper were developed, and him and Patrick Hayden for suggesting significant improvements to previous drafts.

## References

Benjamin, S. (2001) (in preparation).

Deutsch, D. (1985) *Int. J. Theor. Phys.* **24** 1

Deutsch, D. (1997) *The Fabric of Reality,* **Chapter 2, Allen Lane, The Penguin Press, London.**

Deutsch, D. (1999) **A455**, 3129-3197.

Deutsch, D. and Hayden, P. (2000) *Proc. R. Soc. Lond.* **A456**, 1759-1774.

DeWitt, B.S. and Graham, N. (1973) *The Many Worlds Interpretation of Quantum Mechanics.* **Princeton University Press, Princeton NJ.**

Everett, H. (1957) In DeWitt and Graham (1973).

Gottesman, D. (1999) *Group22: Proceedings of the XXII International Colloquium on Group Theoretical Methods in Physics*, **S. P. Corney, R. Delbourgo, and P. D. Jarvis, (eds.) 32-43 (Cambridge, MA, International Press).**

Hartle J.B. (1991) *Phys. Rev.* **D44** 10, 3173.

Page, D.N. and Wootters, W. (1983) *Phys. Rev.* **D27** 12, 2885-2892.

Zurek, W. H. (1981) *Phys. Rev.* **D24** 1516-25.